\documentstyle[prb,tighten,eqsecnum,aps]{revtex}
\begin{document}
\title{\hfill{\large WUB 96-12}\\[3cm]
Critical properties of the one-dimensional spin-$\frac{1}{2}$
antiferromagnetic Heisenberg model in the presence of a uniform field}
\author{A. Fledderjohann,
        C. Gerhardt,
        K.H. M\"utter\footnote{e-mail:muetter@wpts0.physik.uni-wuppertal.de} 
        and A. Schmitt}
\address{Physics Department, University of Wuppertal, 42097 Wuppertal, Germany}
\author{M. Karbach}
\address{Department of Physics, The University of Rhode Island,
Kingston RI 02881, USA}

\date{\today}
\maketitle
%
%
\begin{abstract}
%
%
In the presence of a uniform field the one-dimensional spin-$\frac{1}{2}$
antiferromagnetic Heisenberg model develops zero frequency excitations at
field-dependent 'soft mode' momenta. We determine three types of critical
quantities, which we extract from the finite-size dependence of the lowest
excitation energies, the singularities in the static structure factors and the
infrared singularities in the dynamical structure factors at the soft mode
momenta. We also compare our results with the predictions of conformal
field theory.
\end{abstract}

\draft
%
%
\section{Introduction}
%
%
In this paper we are going to study the zero temperature dynamics of the one
dimensional spin-$\frac{1}{2}$ antiferromagnetic Heisenberg model
\begin{equation}\label{hop}
    H \equiv 2 \sum_{x=1}^N \vec{S}(x) \vec{S}(x+1) - 2 B \sum_{x=1}^N S_3 (x),
\end{equation}
in the presence of a uniform external field $B$. The quantities of interest are
the dynamical structure factors at fixed magnetization $M \equiv S/N$:
\begin{eqnarray}\label{dsf}
	S_a(\omega,p,M,N) = \sum_{n} \delta\biglb( \omega-(E_n-E_s)\bigrb) 
	 |\langle n|S_a(p)|s\rangle|^2, \quad a=3,+,-.
\end{eqnarray} 
They are defined by the transition probabilities $|\langle n|S_a(p)|s\rangle|^2$
from the ground states $|s\rangle\equiv|S,S_3=S\rangle$ in the subspaces with 
total spin $S$ and energy $E_s$ to the excited states $|n\rangle$ with energy 
$E_n$. The transition operators we are concerned with, are the Fourier 
transforms of the single-site spin operators $S_a(x)$:
\begin{equation}\label{fts}
 S_a(p) \equiv \frac{1}{\sqrt N}\sum_{x=1}^{N} e^{i p x} S_a(x), \quad a=3,+,-.
\end{equation}
The structure factors (\ref{dsf}) have been investigated before by M\"uller 
{\it et al.}.\cite{bib1} They performed a complete diagonalization of the 
Hamiltonian (\ref{hop}) on small systems ($N\leq 10$) and analysed the 
spinwave continua by approximately solving the Bethe {\it ansatz} equations 
for the low lying excitations. In particular, they found a lower bound
\begin{equation}\label{lbw}
	\omega \ge |\omega_3(p,M)|,
\end{equation}
\begin{equation}\label{lw3}
	\omega_3(p,M) = 2 D \sin \frac{p}{2} \sin \frac{p-p_3(M)}{2},
\end{equation}
for the excitations contributing to the longitudinal structure factor
$S_3(\omega,p,M)$. The constant $D$ on the right hand side of (\ref{lw3}) is
fixed by the magnetization curve:\cite{bib2}
\begin{equation}\label{mc}
	B(M)=2 D \sin \pi M.
\end{equation}
The lower bound vanishes at $p=0$ and at the field dependent momentum
\begin{equation}\label{fdm}
	p_3(M) = \pi (1-2M),
\end{equation}
signalling the emergence of zero frequency modes ('soft modes') in the spectrum
of excitation energies. The analysis of the spinwave continua relevant for the
transverse structure factors $S_{\pm}(\omega,p,M)$ leads to the
following approximate lower
bounds
\begin{equation}\label{flb}
	\omega \ge \omega_{\pm}(p,M),
\end{equation}
for the excitations produced by the raising and lowering operators $S_{+}(p),
S_{-}(p)$, respectively:
\begin{eqnarray}\label{rlor}
	\omega_{+}(p,M) = 2 D \left[\sin \frac{p}{2} \cos \left(\frac{p}{2}-\pi
	M\right) - \sin \pi M\right] \quad \mbox{ for } \quad p_1(M) \le p \le
	\pi, 
\end{eqnarray}
and
\begin{equation}\label{alor}
	\omega_{-}(p,M) = |\omega_3(\pi-p,M)| 
		\quad \mbox{ for } \quad 0 \le p \le \pi. 
\end{equation}
Both bounds vanish at $p=\pi$ and at $p=p_1(M)=2 \pi M$.  The softmodes at the
field dependent momenta $p_j(M), j=1,3$ produce characteristic structures in 
the momentum dependence of the corresponding static structure
factors.\cite{bib3,bib4} It is the purpose of this paper to analyse 
the singularities in the static
structure factors and the infrared singularities in the dynamical structure
factors (\ref{dsf}) at the softmode momenta. In Sec. II we review our method to
compute the excitation energies and transition probabilities for finite rings
($N \le 36$). The finite-size dependence of the lowest excitation energy at the
soft mode momenta is analysed by 
solving the Bethe {\it ansatz} equations on
large systems ($N \le 2048 $). 
The critical
behavior of the static structure factors at the softmode momenta
$p=p_a(M),\;a=1,3$ and fixed magnetization $M=1/4$ is investigated in Sec. III
based on a numerical computation of the ground state on rings with
$N=12,16,...,32,36$ sites.  In Sec. IV, we demonstrate how infrared
singularities emerge in a finite-size scaling analysis of the dynamical
structure factors in the euclidean time representation. Finally, in
Sec. V we compare our numerical results with the predictions of
conformal field theory.  
%
%
\section{Softmodes in the excitation spectrum.}
%
%
An approximate scheme to determine low lying excitation energies and transition
probabilities has been proposed in Ref.[\onlinecite{bib5}]. It starts from the
recursion algorithm,\cite{bib6} which generates a tridiagonal
matrix. Eigenvalues and eigenvectors of this matrix yield the exact excitation
energies and transition probabilities. There are, however, two sources for
numerical errors in this scheme. The orthogonality of the states produced by 
the recursion algorithm is lost more and more with an increasing number of 
steps, due to rounding errors. Moreover, the iteration has to be truncated 
before the Hilbert space is exhausted.

Nevertheless the method yields good results for the lowest $10$ excitations
-- provided that these contain the dominant part of the spectral
distribution. This condition is satisfied for the excitations in
$S_a(\omega,p,M,N),\, a=3,+$. For $S_-(\omega,p,M,N)$ near the softmode 
momentum $p_1(M)$, however, this is not the case. In Table I we compare the 
low lying excitations for $S_-(\omega,p,M,N),\,M=1/4, p=\pi$ and $p=\pi/2 - 
2\pi/16$ on a ring with $N=16$ sites, as they follow from an exact 
diagonalization (upper part of Table I) and the recursion algorithm (lower 
part of Table I), respectively.


At $p=\pi$, $76.95\%$ of the spectral weight is found in the first excitation.
Energy and relative spectral weight of the first excitation are reproduced
within 13 digits. The following $7$ excitations can be identified term by 
term with decreasing accuracy for the energies and the relative spectral 
weights. 

The situation is different for $p_-=\pi/2-2\pi/16 $, which can be seen in 
the right hand part of Table I. The exact result yields large spectral 
weights -- marked by an asterisk -- for the 1st ($19.55\%$), the 15th 
($18.33 \%$) and the 20th ($13.80 \%$) excitation. The recursion method 
reproduces the energy and spectral weight of the first excitation within 13 
digits. The two other excitations with large spectral weight -- marked by 
an asterisk -- are only in rough agreement with the
exact result. We found, however, that this inaccuracy has no effect on the
dynamical structure factors in the euclidean time representation
(\ref{ietl}). The latter will be investigated in Sec. IV.  In Figs. 1(a),(b),
(c) we present the momentum dependence of the excitation energies in the 
dynamical structure factors $S_a(\omega,p,M=1/4,N=28)$ as they follow from 
the recursion method. The size of the symbols measures the relative spectral 
weight $|\langle n| S_a(p)|s\rangle|^2/S_a(p,M,N)$. The normalization is 
given by the static structure factors:
\begin{equation}\label{gbts}
     S_a(p,M,N) = \int_{\omega_a(p,M,N)}^{\infty} d\omega\; S_a(\omega,p,M,N),
     \quad a=3,+,-.
\end{equation}
There is a strict relation between the static transverse structure factors: 
\begin{equation}\label{tias}
	S_-(p,M,N) = S_+(p,M,N) + 2 M.
\end{equation}
It should be noted that $S_+(p,M,N) \approx 0$ for $p < p_1(M)$ [cf. Fig. 
3(b)], which implies that the absolute spectral weight $|\langle n
|S_+(p)|s\rangle|^2$ is almost zero for $p<p_1(M)$.

The solid curves represent the lower bounds (\ref{lw3}),(\ref{rlor}) and
(\ref{alor}) obtained from the analysis of the spinwave continua.\cite{bib1} 
The emergence of the softmode at $p=p_3(M=1/4)=\pi/2$ in the longitudinal 
case [Fig. 1(a)] is clearly visible. Note, that there are some excitations 
with small spectral weights below the bound (\ref{lw3}) (for $p>3\pi/4$). 
We do not know, whether the spectral weights will survive in the 
thermodynamical limit.

The lowest excitations in the transverse cases [Figs. 1(b) and 1(c)] are 
found at $p=\pi$ and at the field dependent momenta
\begin{equation}\label{tfd}
	p_1^\pm(M) = p_1(M) \pm \frac{2 \pi}{N} .
\end{equation}
We have analysed the finite-size dependence of the lowest excitation energies:
\begin{mathletters}\label{lee}
\begin{eqnarray}
	\omega_3\biglb(p_{3}(M),M,N\bigrb) &=& 
		E\biglb(p=p_s+p_{3}(M),M=S/N,N\bigrb) - E(p_s,M=S/N,N), 
                \label{lee1} \\
	\omega_1\biglb(\pi,M,N\bigrb) &=& 
		E\biglb(p=p_s+\pi,M=(S+1)/N,N\bigrb)-E(p_s,M=S/N,N), 
                \label{lee2} \\
	\omega_{\pm}\biglb(p=p_1^\pm(M),M,N\bigrb) &=& 
		E\biglb(p_s+p_1^\pm(M),M=(S\pm1)/N,N\bigrb) - E(p_s,M=S/N,N), 
                \label{lee3} 
\end{eqnarray} 
\end{mathletters}
$p_s$ denotes the groundstate momentum in the sector with total spin
$S$; $p_s=0$ if $N+2S$ is a multiple of $4$, $p_s=\pi$ otherwise.
The lowest energy eigenvalues $E(p,M,N)$ with momentum $p$ and spin $S$ were
computed on large systems ($N \le 2048 $) by solving the Bethe {\it ansatz}
equations. The extrapolation of the energy differences (\ref{lee}) to 
the thermodynamical limit
\begin{mathletters}\label{tttl}
\begin{eqnarray}
	\lim_{N\to\infty}N\omega_3(p_3(M),M,N)= \Omega_3(M), \quad & &
        \lim_{N\to\infty}N\omega_1(\pi,M,N)= \Omega_1(M),\label{tttl1}\\ 
	\lim_{N\to\infty}N\omega_{\pm}\biglb(p_1^\pm(M),M,N\bigrb)
	        &=& \Omega_1^\pm(M),\label{tttl2} 
\end{eqnarray}
\end{mathletters}
obey the following relations:
\begin{equation}\label{omeg1}
	\Omega_1^\pm(M) =  \Omega_3(M) \pm  \Omega_1(M) .
\end{equation}
Together with the spinwave velocity $v(M)$
\begin{equation}\label{geschw}
	 2 \pi v(M)=\lim_{N  \to \infty} N[E(p_s+2\pi/N,M,N)-E(p_s,M,N)],
\end{equation}
they define the scaled energy gaps:
\begin{mathletters}\label{teta}
\begin{eqnarray}
	2 \theta_a(M) &=& \frac{\Omega_a(M)}{\pi v(M)}, \quad a=3,1, 
        \label{tetaa} \\
	2 \theta_1^\pm(M) &=& \frac{\Omega_1^\pm(M)}{\pi v(M)} 
	= 2 [\theta_3(M)\pm\theta_1(M)]	\label{tetein}. 
\end{eqnarray}
\end{mathletters}
The $M$-dependence of the quantities $\theta_a(M),
\; a=3,1$, is shown in Fig. 2. It turns out that
\begin{equation}\label{quot}
	2 \theta_1(M) = \frac{1}{2 \theta_3(M)}
\end{equation}
in accord with the analytical result of Bogoliubov, Izergin and
Korepin \cite{bib7}.
In the limit $M \rightarrow 1/2$ one finds $2 \theta_3(M) = 1+2M$.\cite{bik} 
The dotted line in Fig. 2 near $M = 0$ indicates the logarithmic singularity
\begin{equation}\label{celim2}
	2 \theta_3(M) \stackrel{M\to0}{\longrightarrow} 
		1+\left(\ln\frac{1}{M^2}\right)^{-1},
\end{equation}
which was obtained by Bogoliubov, Izergin and Korepin \cite{bik} by a
perturbative approach to the Bethe {\it ansatz} equations.
%
%
\section{Critical behavior of the static structure factors at the softmode 
momenta}
%
%
The static structure factors of the antiferromagnetic Heisenberg model in the
presence of a magnetic field have been investigated in a previous numerical
study on systems up to $N=28$.\cite{bib4} Meanwhile we have extended the system
size to $N=32$ and $N=36$ at fixed magnetization $M=1/4$. We find the following
features:
\begin{enumerate}
\item The transverse structure factor at momentum $p=\pi$ diverges for
	$N \to \infty$. A power law fit
	\begin{equation}\label{powlaw}
       	S_1(\pi,M,N) \stackrel{N\to\infty}{\longrightarrow} 0.503 
        N^{1-\eta_1(M)}, 
	\end{equation}	
	to the finite system results for $N=36,32,28$ leads to the value
	$\eta_1(M=1/4)= 0.65$ for the critical exponent. The same exponent
	governs the approach to the singularity in the momentum $p$:
	\begin{equation}\label{mompe}
	  S_1(p,M,\infty) \stackrel{p\to\pi}{\longrightarrow}
	  0.316 \left(1-\frac{p}{\pi}\right)^{\eta_1(M)-1},
	\end{equation}
	The finite-size dependence (\ref{powlaw}) is shown in Fig. 3(a).
        The momentum dependence can be seen in Fig. 3(b) where we have
        plotted $S_1(p=\pi,M=\frac{1}{4},N)$ versus $(1-\frac{p}
        {\pi})^{\eta_1(M)-1}$ 
        using the critical exponent determined in Fig. 3(a).
\item The approach to the field dependent soft mode $p_1(M)=2\pi M$
	in the transverse structure factor is shown in the upper left $[p\to
	p_1(M)-0]$ and the lower right $[p \to p_1(M)+0]$ insets of Fig. 3(b). 
        The numerical data behave as
	\begin{equation}\label{powf2}
		S_1\biglb(p\to p_1(M)\pm 0,M,\infty\bigrb)\sim 
		\left | 1 - \frac{p}{p_1(M)}\right | ^{\eta_1^\pm(M) -1}.
                \nonumber
	\end{equation}
	if the critical exponents are chosen to be $\eta_1^+(M=1/4)=2.17$ ,
	$\eta_1^-(M=1/4)=0.8 ... 1.2$.  The uncertainty in
	$\eta_1^-(M=1/4)$ reflects an instability in the fit to the numerical
	data. Note, that the right hand side of (\ref{powf2}) diverges for
	$\eta_1^-(M=1/4)<1$ but converges for $\eta_1^-(M=1/4)>1$. An
	unambiguous determination of $\eta_1^-(M=1/4)$ demands for much larger
	systems than $N=36$.

\item The finite-size dependence of the longitudinal structure factors
	at $p=p_3(M)$
	\begin{equation}\label{powf3l}
	 	S_3\biglb(p_3(M),M,N\bigrb) \stackrel{N\to\infty}
                {\longrightarrow} -0.124 N^{1-\eta_3(M)} + 0.308,
	\end{equation}
	is shown in Fig. 4(a) for $M=1/4,\; p=p_3(M)=\pi/2$.
	A power law fit to the finite system results with $N=36,32,28$ yields:
	$\eta_3(M=1/4)=1.51$. The same exponent governs the approach to the
	singularity from the left:
	\begin{equation}\label{fleft}
		S_3\biglb(p\to p_3(M)-0,M,N\bigrb) 
                \stackrel{N\to\infty}{\longrightarrow}
		-0.312\left(1-\frac{p}{p_3(M)}\right)^{\eta_3(M)-1} + 0.322,
	\end{equation}
	as is demonstrated in Fig. 4(b). It is not so easy to decide, whether a
	different exponent is needed to describe the approach to the 
        singularity from the right. In the inset of Fig. 4(b) we plotted the 
        approach from the
	right versus $|1-p/p_3(M)|^{\eta_3(M=1/4)-1}$.

	The Fourier transform of the singularities in the static structure 
        factors determines the large distance behavior of the corresponding 
        spin spin correlators:
	\begin{mathletters}\label{ldbe}
	\begin{eqnarray}\label{ldbe1}
		\langle s|S_1(0) S_1(x)|s\rangle \;&& 
                \stackrel{x\to\infty}{\longrightarrow} \; 
 			\cos(\pi x) \frac{A_1(M)}{x^{\eta_1(M)}} 
		+ \cos[p_1(M)x]
			\left(\frac{A_1^+(M)}{x^{\eta_1^+(M)}} + 
				  \frac{A_1^-(M)}{x^{\eta_1^-(M)}}\right), \\
		\langle s|S_3(0)S_3(x)|s\rangle-\langle
                s|S_3(0)|s\rangle^2 \; 
		&&	\stackrel{x\to\infty}{\longrightarrow} \; 
		 \cos [p_3(M) x] \frac{A_3(M)}{x^{\eta_3(M)}}. \label{ldbe2}
	\end{eqnarray}
	\end{mathletters}
	Conformal field theory \cite{bib9} predicts a relation between the 
        critical exponents $\eta(M)$ in (\ref{ldbe}) and the 
        scaled energy gaps (\ref{teta}):\cite{bib7,bib8}
	\begin{mathletters}\label{reled}
	\begin{eqnarray}
		2 \theta_a(M)    &=& \eta_a(M), \quad a=3,1, \\
		2 \theta_1^\pm(M)&=& \eta_1^\pm(M).
	\end{eqnarray}
	\end{mathletters}
	A derivation of (\ref{reled}) is presented in appendix A.  A 
        comparison of the left and right hand sides of (\ref{reled}) is 
        presented in Table II. 
\end{enumerate}
%
%
\section{Finite-size scaling analysis of the infrared singularities}
%
%
The euclidean time representation 
\begin{eqnarray}\label{ietl}
	S_a(\tau,p,M,N) = \int^{\infty}_{\omega_a(p,M,N)} d\omega \;
	e^{-\omega\tau}S_a(\omega,p,M,N), \quad a=3,+,-,
\end{eqnarray}
is most suited to study finite-size effects in the dynamical structure factors
(\ref{dsf}).  The singularities in the static structure factors
$S_a(\tau=0,p,M,N)$ at the softmode momenta originate from the infrared
singularities in the dynamical structure factors.  In the combined limit
\begin{equation}\label{ctcl}
	\tau \to \infty, \qquad N\to\infty,  
\end{equation}
- keeping fixed the 'scaling' variables -
\begin{equation}\label{scvar}
	z_a(p,M)\equiv \tau \omega_a(p,M,N), \quad a=3,+,- ,
\end{equation}
the low frequency part at the softmode momenta $p=\pi,p=p_1(M)\pm
2\pi/N,p=p_3(M)$ is projected out. We therefore expect to see here directly
signatures for the infrared singularities. Let us assume that the emergence of
the infrared singularities on finite systems can be described by a finite-size
scaling ansatz:
\begin{equation}\label{fsans}
	S_a(\omega,p,M,N) = \omega^{-2\alpha_a(p,M)} 
	g_a\biglb(\omega/\omega_a(p,M,N),n_a(p,M,N)\bigrb),\quad a=3,+,-.
\end{equation}
The scaling functions $g_a$ are supposed to depend only on the scaled 
excitation energies $\omega/\omega_a(p,M)$ and the variable
\begin{equation}\label{hvar}
	n_a(p,M,N) = [p-p_a(M)]N/(2\pi),
\end{equation}
which describes the approach to the softmode momenta.  The ansatz (\ref{fsans})
induces the following finite-size scaling behavior of the euclidean time
representation (\ref{ietl}) in the combined limit (\ref{ctcl}) and
(\ref{scvar}):
\begin{equation}\label{bitl}
	\tau^{1-2\alpha_a(p,M)} S_a(\tau,p,M,N) =
	G_a\biglb(z_a(p,M),n_a(p,M,N)\bigrb) \exp[-z_a(p,M)].
\end{equation}
The two scaling functions on the right hand sides of equations
(\ref{fsans}) and (\ref{bitl}) are related via:
\begin{equation}
G(z,n) = z^{1-2 \alpha} \int_1^{\infty} dx \; e^{-(x-1)z}
g(x,n) .
\nonumber
\end{equation}
Based on our numerical results for $S_a(\tau,p,M,N)$ at
$M=\frac{1}{4}$, $a=3,+$, $N=16,20,...,36$ and $a=-$,  
$N=16,20,...,32$
at the softmode momenta we will now
test the validity of the finite-size scaling ansatz (\ref{bitl}).

Let us start with the longitudinal structure factor at the softmode
$p=p_3(M=1/4) =\pi/2$. In this case the variable (\ref{hvar}) is
$n_3(p=\pi/2,M=1/4) = 0$. The left hand side of (\ref{bitl}) versus the 
scaling variable $z_3(p=\pi/2, M=1/4)$ is shown in Fig. 5(a) for the 
following values of $\alpha_3(p=\pi/2,M=1/4)=0.22, 0.23, 0.234$. For $z_3 
\ge 0.4 $ (inset of Fig. 5(a)) the finite system results coincide best if
\begin{equation}\label{zwodrei}
	\alpha_3(p=\pi/2, M=1/4) = 0.23 .
\end{equation}
Therefore, this is the expected critical exponent for the infrared singularity
in the longitudinal structure factor. Deviations from this value for 
$\alpha_3$ on the left hand side of (\ref{bitl}) obviously lead to a violation 
of finite size scaling. It is remarkable to note that finite-size scaling 
[with the exponent $\alpha_3(p=\pi/2, M=1/4) = 0.23$] persists for all values 
$z_3 \ge 0.4$. In the limit $z_3 \to \infty$ the first excitation alone 
survives and we can conclude for the finite-size dependence of the transition 
probability:
\begin{equation}\label{trprob}
	|\langle n=1|S_3(p=\pi/2)|s\rangle|^2 \stackrel{N\to\infty}
        {\longrightarrow}
	N^{2\alpha_3-1}.
\end{equation}
In other words, the critical exponent $\alpha_3$ for the infrared singularity
can by read off the finite-size dependence of the transition probability for 
the first excitation. Indeed this feature is predicted by conformal field
theory.\cite{bib7} (cf. (A9) in appendix A)

Next we turn to the infrared singularities of the transverse structure factors
$S_{\pm}(\omega,p=\pi,M=1/4)$. As can be seen from Fig. 5(b), finite-size
scaling is found for the following choice of the critical exponents:
\begin{mathletters}\label{choice}
\begin{eqnarray}
	\alpha_+(p=\pi,M=1/4) &=& 0.69,\\
	\alpha_-(p=\pi,M=1/4) &=& 0.66 .
\end{eqnarray}
\end{mathletters}
In contrast to the longitudinal case, finite-size scaling can be observed here
for all values of the scaling variables $z_+, z_-$. 

Finally we present in Figs. 6(a) and 6(b) the tests of finite-size scaling for
the transverse structure factors $S_{\pm}(\tau, p=\pi/2 \pm 2\pi/N, M=1/4, N)$
if we approach the field dependent soft mode $p_1(M=1/4)=\pi/2$ from the left
$(p=\pi/2-2\pi/N)$ and from the right $(p=\pi/2+2\pi/N)$, respectively. The
critical exponents are found to be
\begin{mathletters}\label{rtce}
\begin{eqnarray}
	\alpha_+(p=\pi/2+2\pi/N,M=1/4) &=& -0.20, \\
	\alpha_-(p=\pi/2-2\pi/N,M=1/4) &=& -0.05.
\end{eqnarray}
\end{mathletters}
Finite-size scaling works quite well for $S_+$ for large and small values of 
the scaling variable $z_+$ as can be seen from the inset in Fig. 6(a). This 
is not the case for $S_-$. Here finite-size scaling breaks down for small 
values of $z_-$ as is demonstrated in the inset of Fig. 6(b). The critical 
exponent $\alpha_-(p=\pi/2-2\pi/N,M=1/4)=-0.05$ results from the finite size
scaling analysis for large values of $z_-$, where the transition
probability for the first excitation is projected out and has the
following finite size dependence:
\begin{equation}\label{fexa}
	|\langle n=1|S_-(p=\pi/2-2\pi/N)|s\rangle|^2
		\stackrel{N\to\infty}{\longrightarrow} N^{2\alpha_- -1}.
\end{equation}
%
%
\section{Discussion and conclusions.}
%
%
In the presence of a uniform field, the one-dimensional antiferromagnetic
Heisenberg model is critical in the following sense:
The excitation spectrum is gapless at the momenta $p=0, \; p=\pi, \;
p=p_3(M)=\pi(1-2M) $ and $p=p_1(M)=\pi \cdot 2 M$. 
In this paper we have tried to answer the following question: Is
conformal field theory applicable to describe the low energy
excitations at these momenta ? To answer this question we have determined:
\begin{enumerate}
\item the scaled energy gaps $2 \theta(M)$, defined through (\ref{lee})-
(\ref{teta})

\item the critical exponents $\eta(M)$ for the singularities
        (\ref{mompe}), (\ref{powf2}) and (\ref{fleft})
	in the static structure factors

\item the exponents $\alpha(M)$ for the infrared singularities
        (\ref{fsans}) 
        in the dynamical structure factors
\end{enumerate}
A compilation of the various critical quantities for $M=1/4$ is given in 
Table II.


The predictions of conformal field theory are reviewed in appendix A.
In particular the following relation is expected to hold:
\begin{equation}\label{ceis}
	2 \theta(M) = \eta(M) = 2[1-\alpha(p,M)].
\end{equation}
Looking at Table II we find:
\begin{itemize}
\item[(a)] The critical quantities $2 \theta_3(M=1/4),\;
	\eta_3(M=1/4)$ and $2-2\alpha_3(p=\pi/2,M=1/4)$ agree within the 
        numerical uncertainty.  Moreover, the critical exponent
	$\alpha_3(p=\pi/2, M=1/4)$ also governs the finite-size dependence 
        of the 	transition probability for the lowest excitation 
        (\ref{trprob}).         We therefore conclude, that the excitations 
        in the longitudinal
	structure factors at the softmode $p_3(M)=\pi(1-2M)$ are correctly
	described by conformal field theory.

\item[(b)] The critical quantities $2 \theta_1(M=1/4),
	\eta_1(M=1/4), 2-2\alpha_+(p=\pi,M=1/4), 2-2\alpha_-(p=\pi,M=1/4) $ 
        agree within numerical uncertainties.  
        In both cases the finite-size dependence of the
	transition probability for the lowest excitation is in accord with the
	prediction of conformal field theory.

\item[(c)] The critical quantities $2\theta_1^+(M=1/4)$ and
	$\eta_1^+(M=1/4)$ agree within numerical uncertainties and deviate
	by about $15 \%$ from the exponent $2[1-\alpha_+
        (p=\pi/2+2\pi/N,M=1/4)]$.

\item[(d)] The scaled energy gap $2\theta_1^-(M=1/4)$ agrees with the
        critical exponent $\eta_1^-(M=1/4)$ - within the large
        numerical uncertainty -  but
	strongly deviates by more than a factor of 2 from the exponent
	$2[1-\alpha_-(\pi/2-1/(2N),M=1/4)]$, which we extracted from the 
        finite-size 	scaling analysis of the infrared singularity in 
        the transverse structure factor $S_-$ at the softmode 
        $p=p_1(M)-2\pi/N,\;M=1/4$. It was demonstrated
	in Fig. 6(b) that finite-size scaling only works for large values of 
        the variable $z_-$, where the first excitation alone contributes. 
        Therefore, the
	exponent $2[1-\alpha(\pi/2-2\pi/N,M=1/4)]$ is fixed by the
	finite-size behavior (\ref{fexa}) of the transition probability for the
	first excitation. The exponent is definitely different from the 
	scaled energy gap $2\theta_1^-(M=1/4)$. 
\end{itemize}
It is worthwhile to note that in the cases (a), (b) and (c), where we find
agreement of our numerical results with the prediction (\ref{ceis}) of 
conformal field theory the spectral weight of the excitations is concentrated 
at low frequencies. This can be seen directly for the case (b) ($p=\pi$) in 
the left hand part of Table I. In contrast, the right hand part of Table I 
shows the widespread distribution of the spectral weight for case (d). Here 
we were not able to establish the identity (\ref{ceis}).
%
%
\section*{Acknowledgement}
%
%
We are indebted to Prof. K. Fabricius, who made us available the exact
numerical results in the upper part of Table I.
We thank Prof. G. M\"uller for helpful comments on this paper.
M. Karbach gratefully acknowledges support by the Max Kade Foundation.
C. Gerhardt was supported by the Graduiertenkolleg 'Feldtheoretische
und numerische Methoden in der Elementarteilchen Physik und
Statistischen Physik'.
%
%
\begin{appendix}
\appendix
%
%
%
%
\section{Critical exponents in conformal field theory} 
%
%
In the absence of a magnetic field the spin-$\frac{1}{2}$ Heisenberg model is
known to be conformal invariant.  Switching on the magnetic field rotational
invariance is broken explicitly. Nevertheless the system remains
gapless. Let us assume that 
the low energy physics of the model is still governed by
conformal field theory.  Then the dominant
contribution to the long distance asymptotics of the zero-temperature dynamical
correlation functions in the infinite $x-t$ plane is correctly
described as \cite{bik}
\begin{eqnarray}\label{CQTHcorr}
	\langle s|S_a(0,0)S_a(x,t)|s\rangle -\langle s|S_a(0,0)|s\rangle^2 
	&=& e^{ixp_a(M)}
	\frac{A_a(M)}{[x+v(M)t]^{2\Delta_a(M)}[x-v(M)t]^{2\bar{\Delta}_a(M)}}.
\end{eqnarray}
$v(M)$ is the spin wave velocity defined in (\ref{geschw}),   
$\Delta_a(M)$ and $\bar{\Delta}_a(M)$ are
the conformal dimensions of the operator $S_a(x,t)$.  The dynamical structure
factor $S_a(\omega,p)$ is just the Fourier transform of (\ref{CQTHcorr}) with 
an appropriate regularization.  The latter can be achieved by giving an
infinitesimal imaginary part to the spinwave velocity $v(M)$.  Standard 
methods yield
\begin{equation}\label{CQTHstf}
	S_a(\omega,p) \sim 
	\left\{\omega \mp v(M)[p-p_a(M)]\right\}^{2\Delta_a(M)+
        2\bar{\Delta}_a(M)-2},
\end{equation}
near the singularities
\begin{eqnarray}
\omega \approx \pm v(M)[p-p_a(M)].
\end{eqnarray}
Equation (\ref{CQTHstf}) is obtained if we first consider the case
$\Delta_a(M)+\bar{\Delta}_a(M) > 1/2$ and then continue analytically. A
conformal transformation to a strip geometry of width $N$ tells us how the
conformal dimensions $\Delta_a(M)$ and $\bar{\Delta}_a(M)$ are related to the
energy and momentum of the lowest excitation $|1\rangle$ provided that the
transition matrix element $\langle s|S_a(0,0)|1\rangle$ does not vanish:
\begin{mathletters}
\begin{eqnarray}
	2\Delta_a(M)      &=& \theta_a(M)+n_a, \\
	2\bar{\Delta}_a(M)&=& \theta_a(M)-n_a, 
\end{eqnarray}
\end{mathletters}
where
\begin{eqnarray}
	n_a = [p-p_a(M)] \frac{N}{2\pi}.
\end{eqnarray}
Therefore we conclude that the infrared singularity of the dynamical structure
factor
\begin{eqnarray}
	S_a(\omega,p) \sim 
	\frac{1}{\left\{\omega \pm v(M)[p-p_a(M)]\right\}^{2\alpha_a(M)}},
\end{eqnarray}
is independent of $n_a$:
\begin{eqnarray}
	\alpha_a(M) = 1-\theta_a(M).
\end{eqnarray}
The critical exponent $\eta_a(M)$ can be read off directly from 
(\ref{CQTHcorr}):
\begin{equation}
	\eta_a(M) = 2\Delta_a(M)+2\bar{\Delta}_a(M) = 2\theta_a(M) .
\end{equation}
In (\ref{CQTHcorr}) it is assumed that the coefficient $A_a(M)$ is 
nonvanishing. From the conformal transformation to the strip geometry a 
relation between $A_a(M)$ and the transition matrix element can be derived:
\begin{eqnarray}
	A_a(M)=\lim_{N \rightarrow\infty}
		\left[2\left(\frac{N}{\pi}\right)^{2\theta_a(M)}e^{i\pi n_a}
	|\langle s|S_a(x,0)|1\rangle|^2 \right].
\end{eqnarray}
Therefore, the matrix element is expected to scale as
\begin{equation}
	|\langle s|S_a(x,0)|1\rangle|^2 \sim N^{2\alpha_a(M)-2} .
\end{equation}
If a finite-size analysis of these critical exponents reveals that
\begin{equation}
	\theta_a(M) < 1 - \alpha_a(M),
\end{equation}
the coefficient $A_a(M)$ vanishes. In this case the expression (\ref{CQTHcorr})
does not represent the dominant contribution to the dynamical structure factor.
\end{appendix}
%
%

%
%
\newpage
\centerline{\bf Figure Captions}
%
%
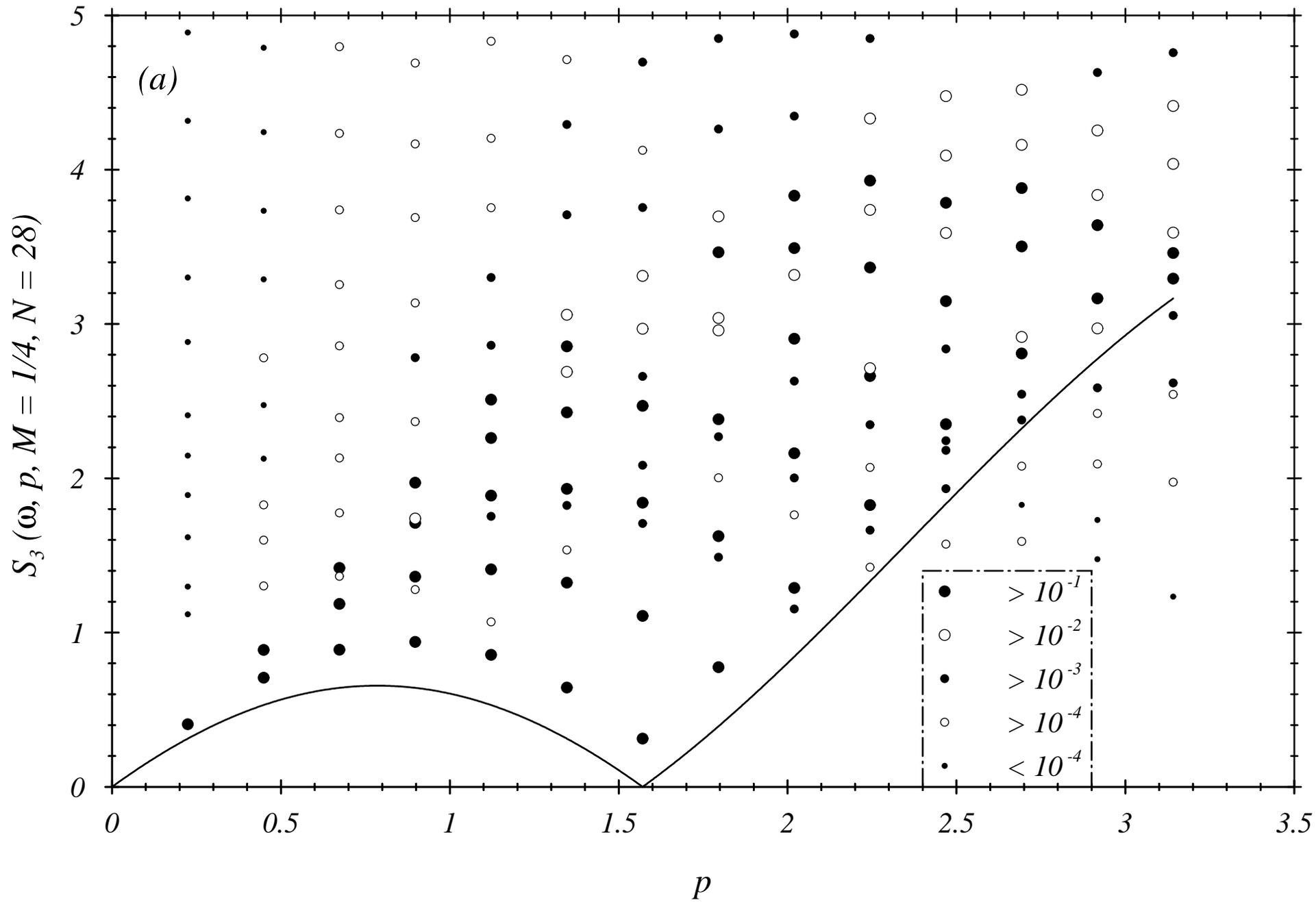
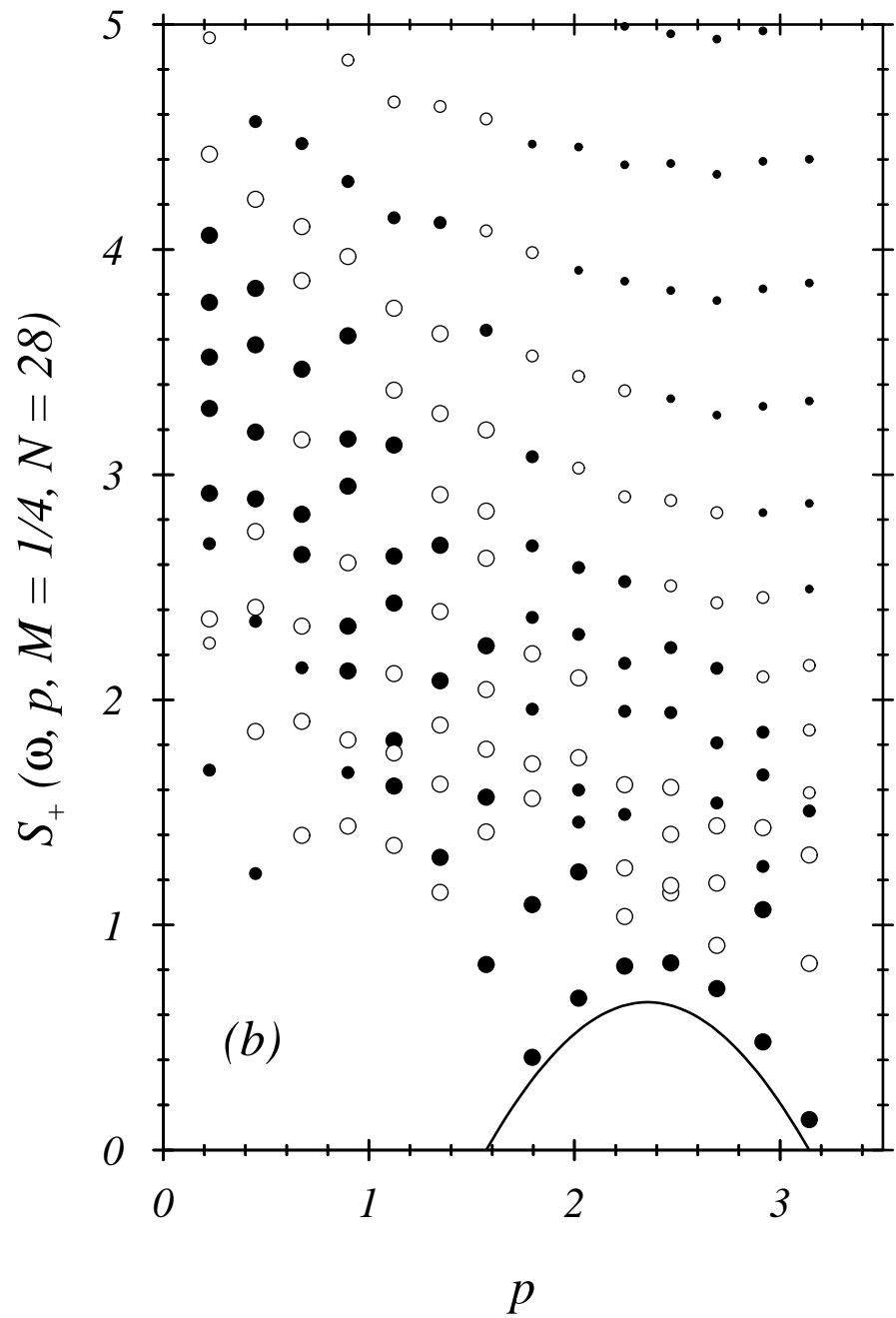
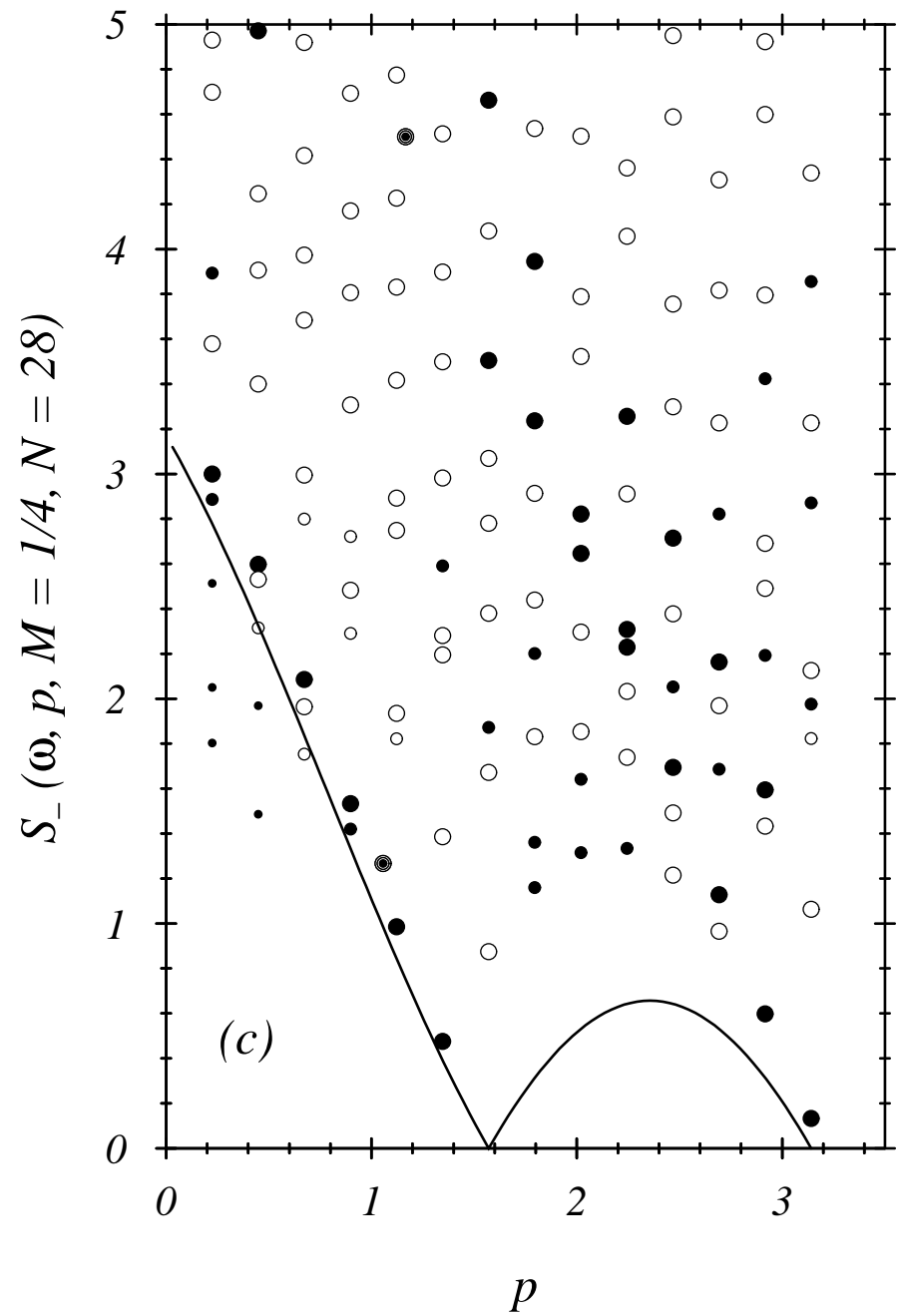
\begin{figure}
\caption[1]{
 	Momentum dependence of the excitation energies in the dynamical 
        structure factors at $M=1/4$: (a) $S_3(\omega,p,M=1/4,N=28)$, (b)
	$S_+(\omega,p,M=1/4,N=28)$, (c) $S_-(\omega,p,M=1/4,N=28)$.  The 
        relative 	spectral weight is characterized by the different 
        symbols.}
\end{figure}

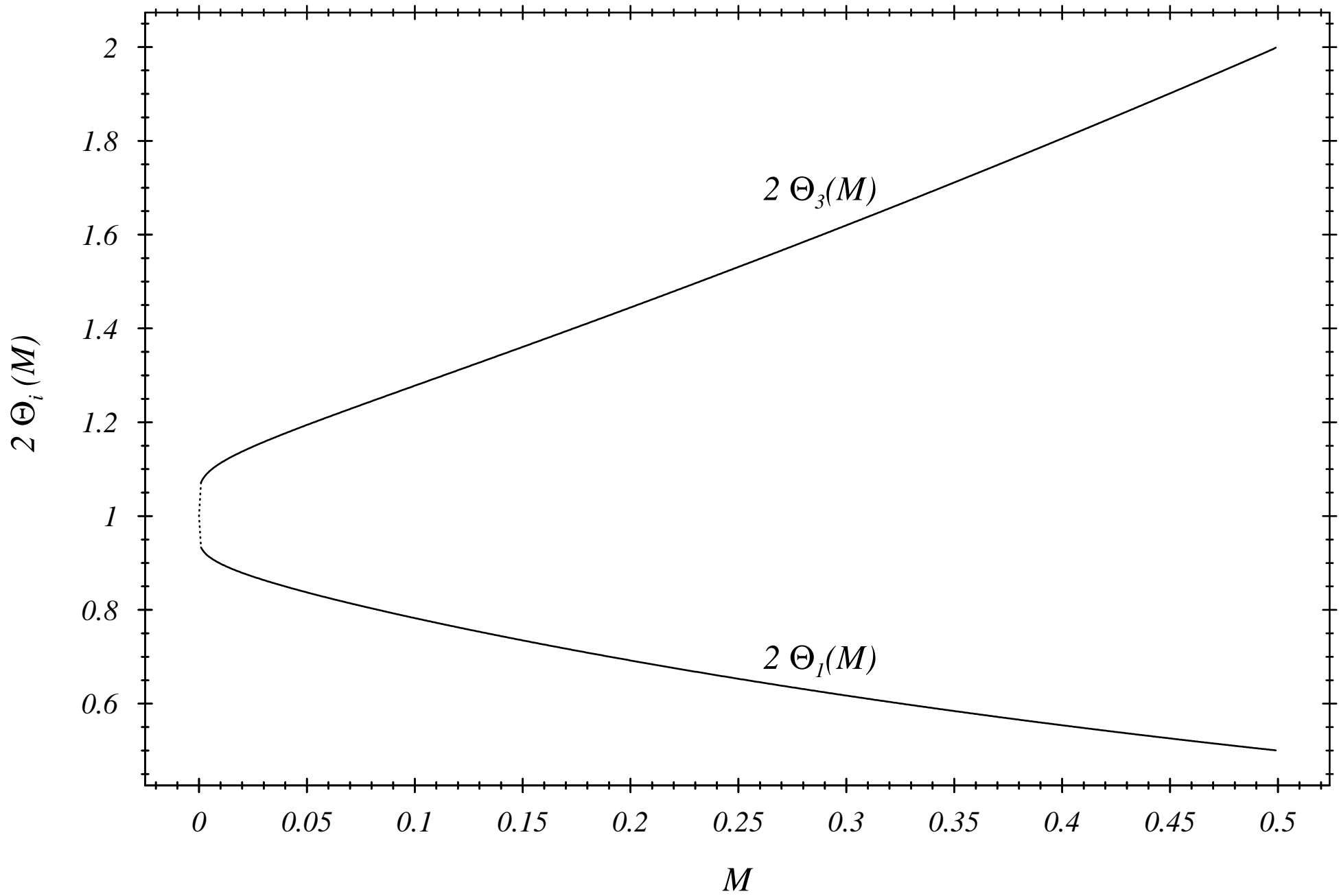
\begin{figure}
\caption[2]{ 
	The dependence of the scaled energy gaps $\theta_1(M)$,
	$\theta_3(M)$ on the magnetization $M$.}
\end{figure}

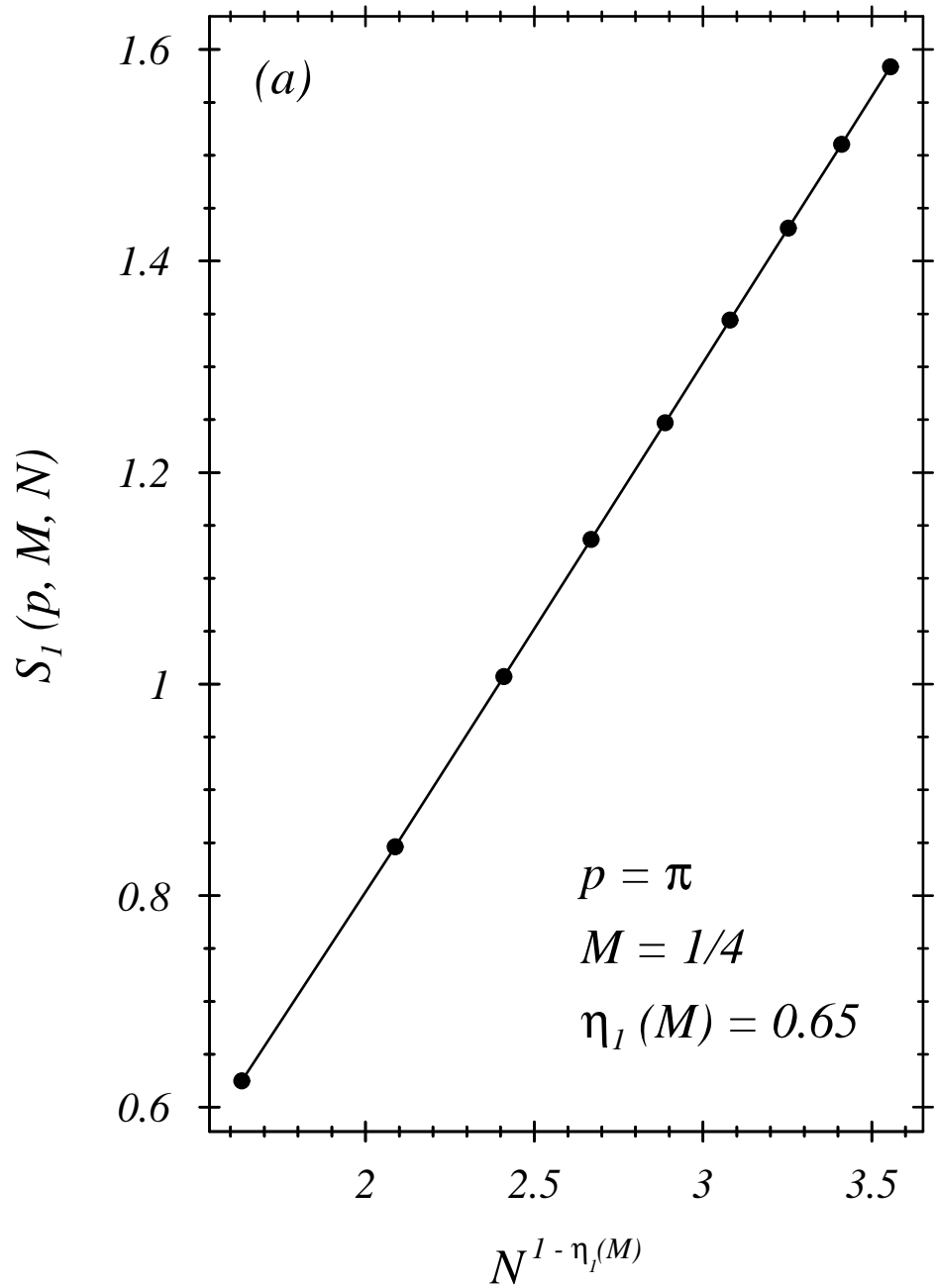
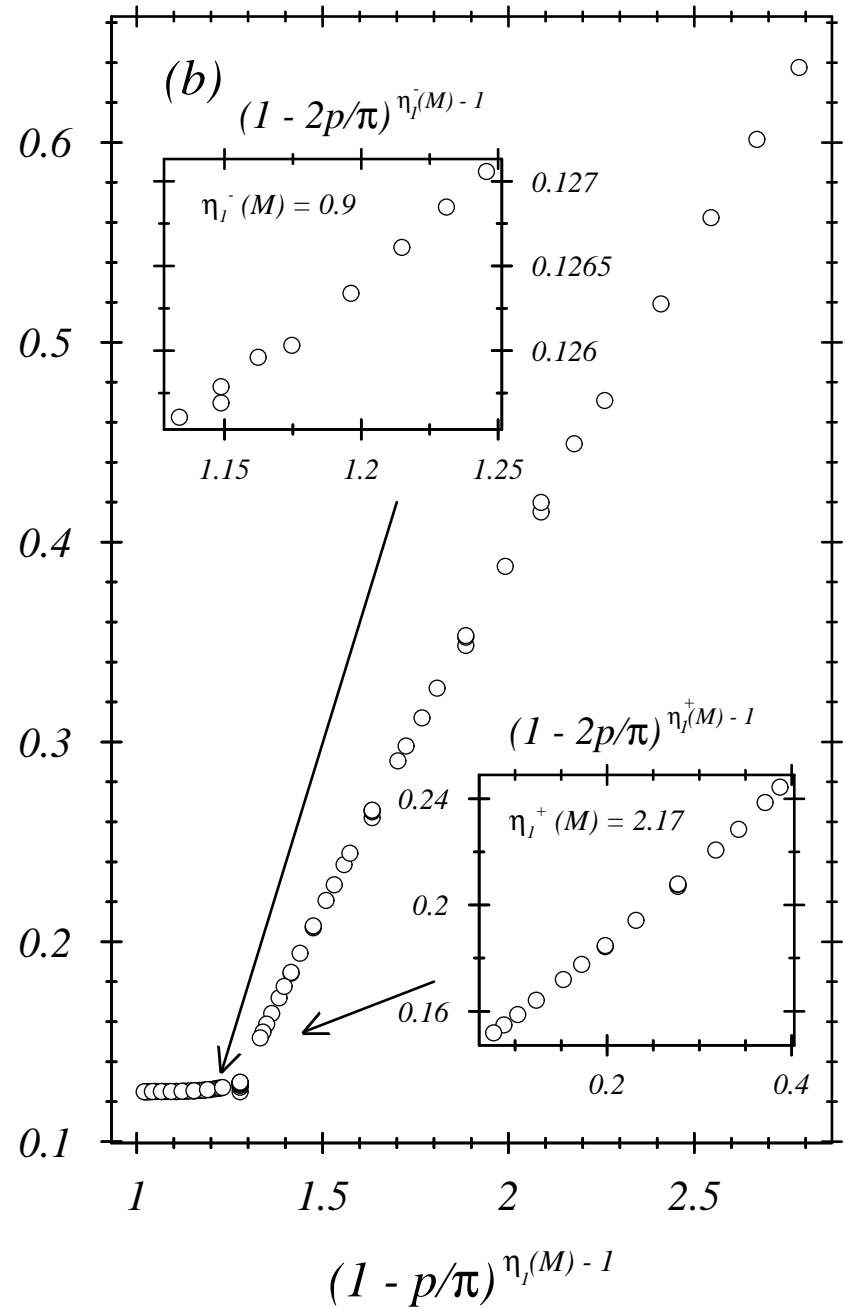
\begin{figure}
\caption[3]{
 	The transverse static structure factor at $M=1/4$: (a) finite-size 
        behavior     at $p=\pi$; (b) the momentum dependence 
        $(1-p/\pi)^{\eta_1(M)-1}$ for $p \to \pi$; $(1-2p/\pi)^
        {\eta_1^-(M)-1}$ for $p \to \pi/2-0$ (inset upper left) and 
        $|1-2p/\pi|^{\eta_1^+(M)-1}$ for $p \to \pi/2+0$ (inset lower right)}.
\end{figure}

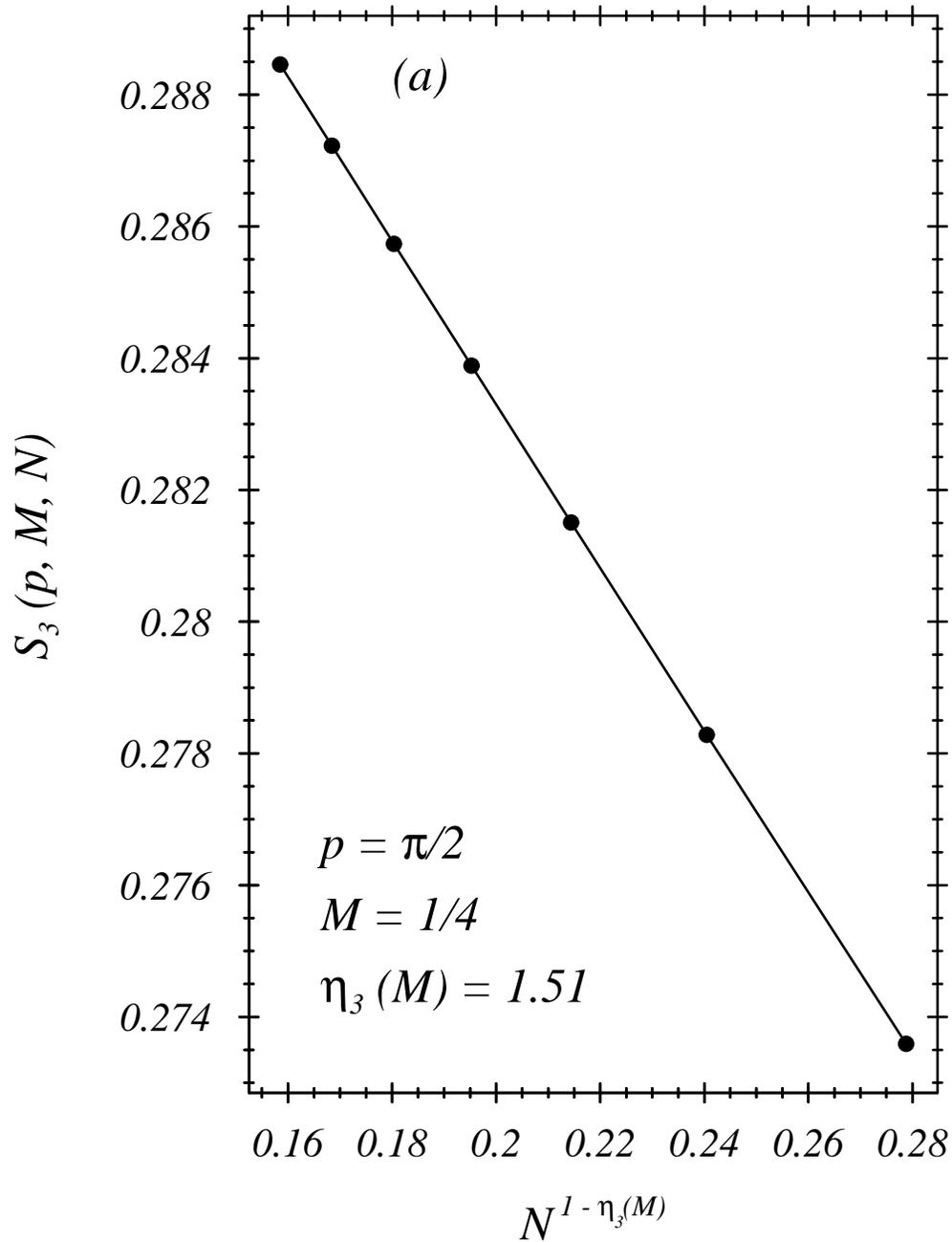
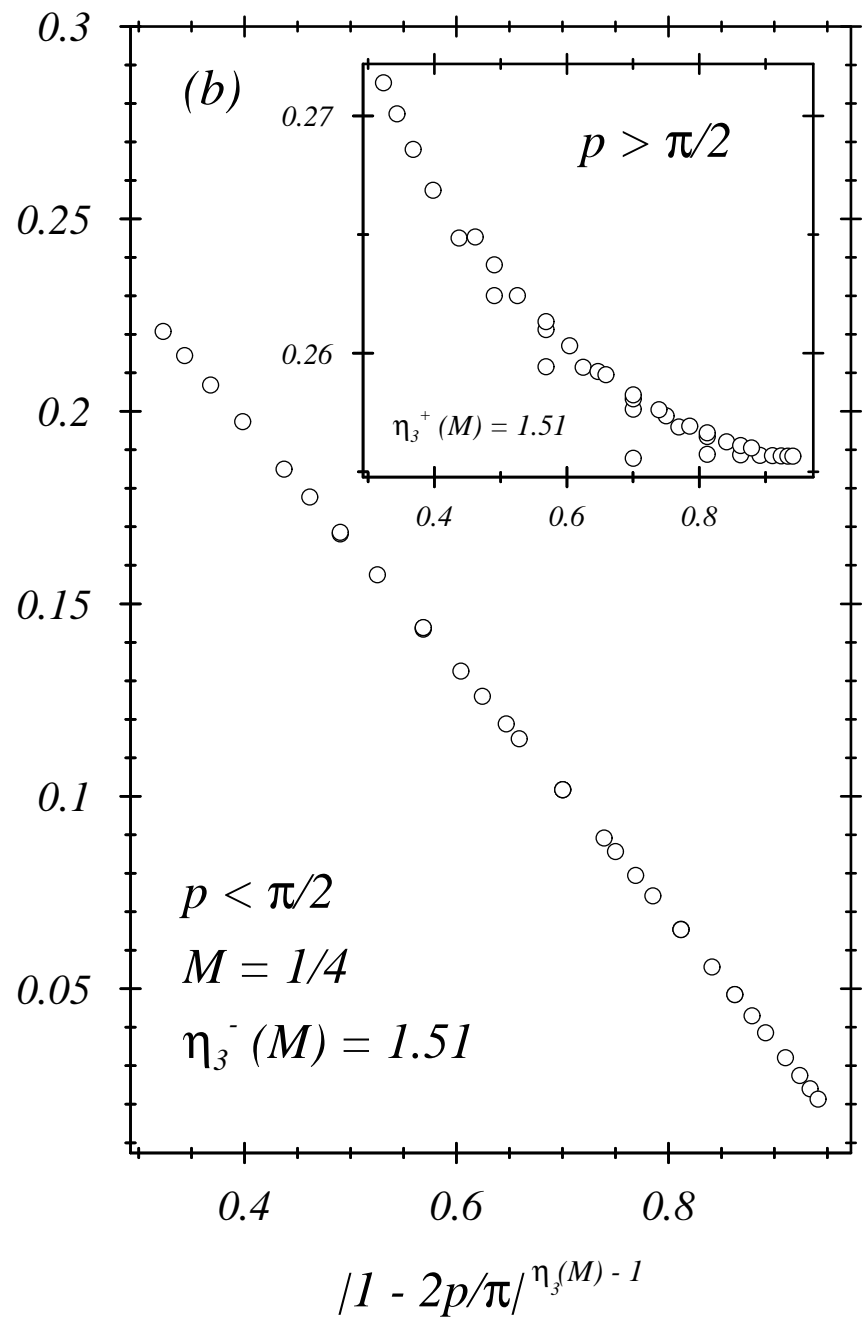
\begin{figure}
\caption[4]{
	The longitudinal static structure factor at $M=1/4$; (a) finite-size 
        behavior
	at $p=p_3(M)=\pi/2$ (b) the momentum dependence: $ \mid 1-2 p/\pi \mid
	^{\eta_3(M)-1}$ for $p<\pi/2$ and $p>\pi/2$ (inset), respectively.}
\end{figure}

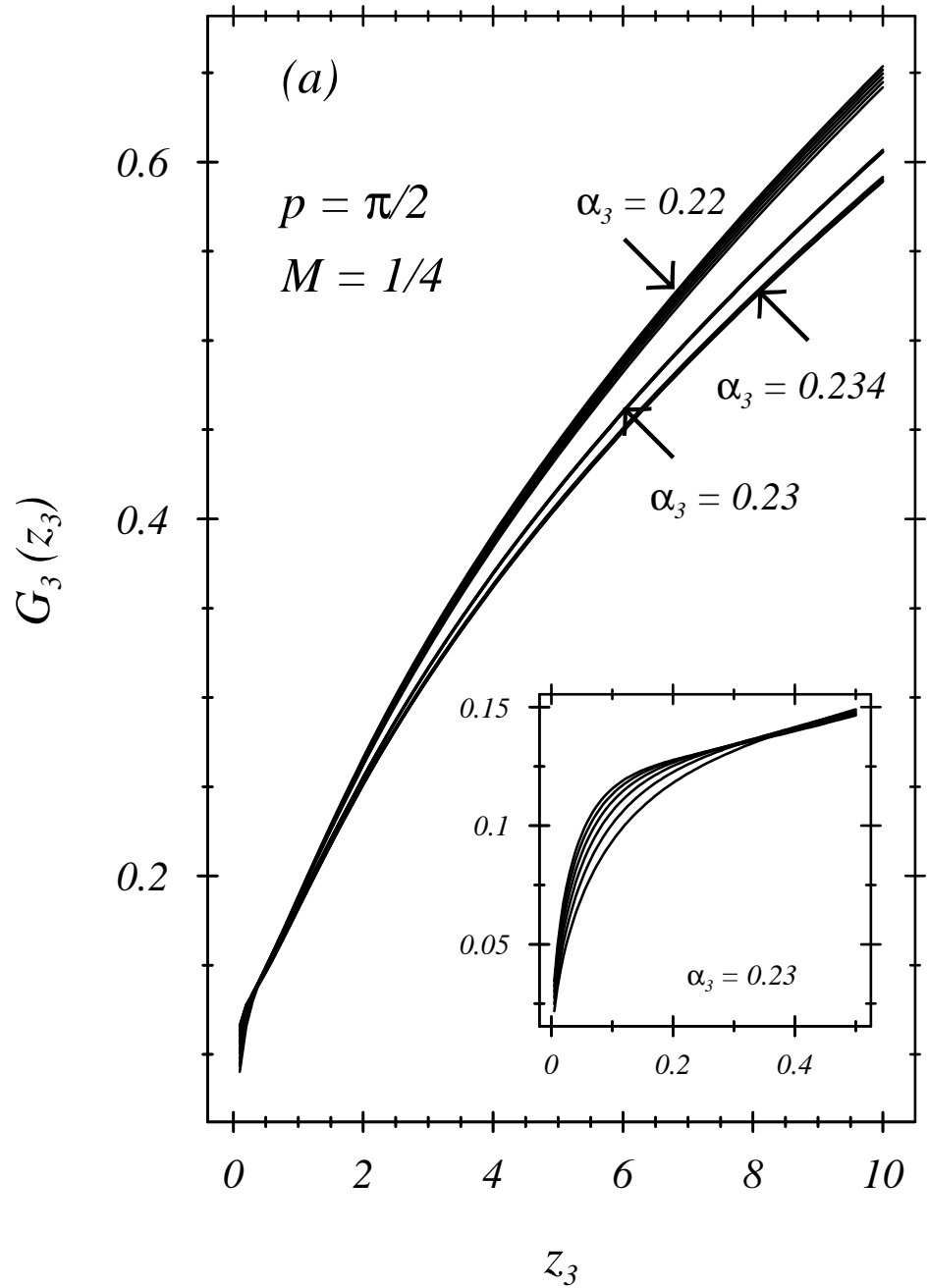
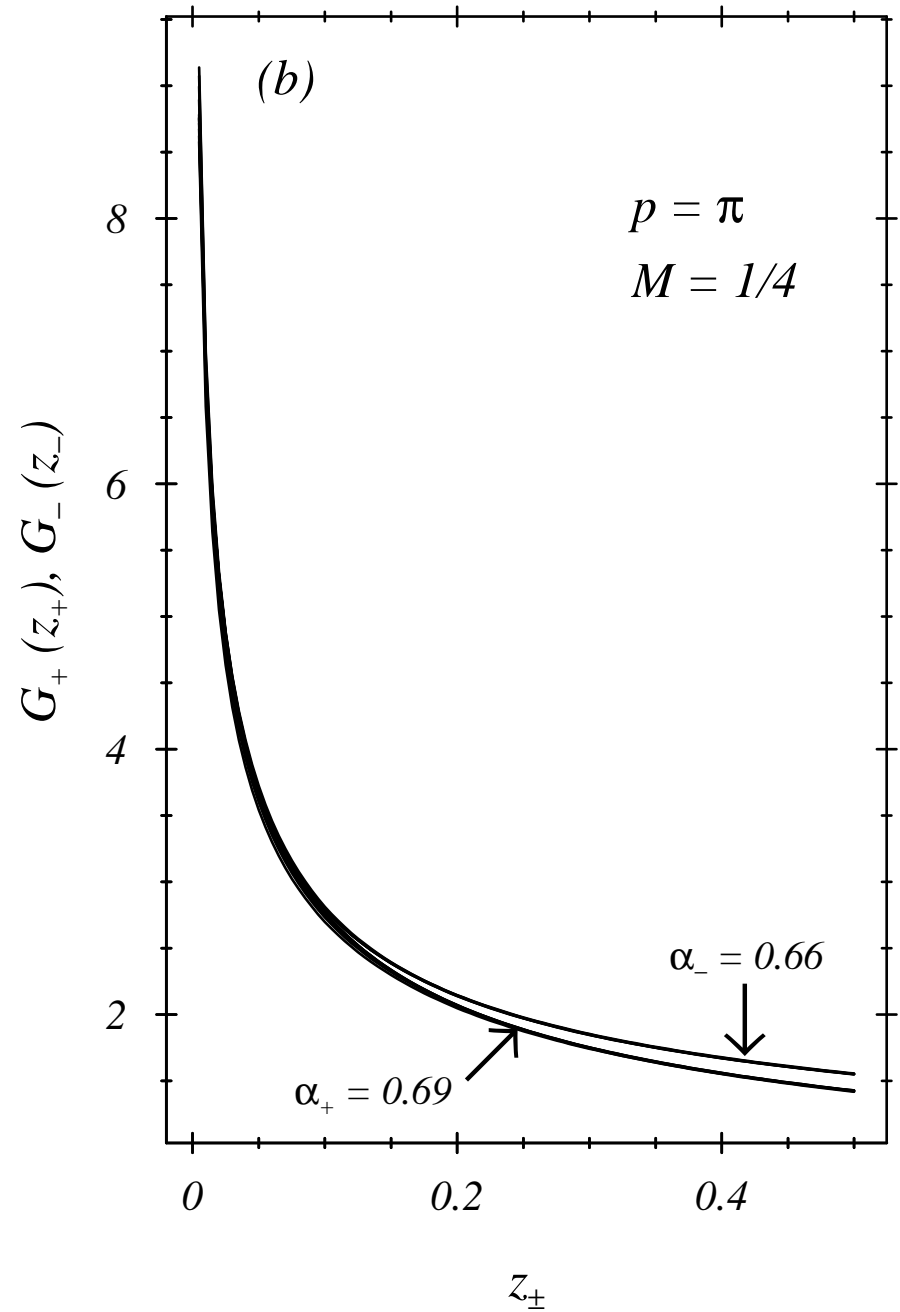
\begin{figure}
\caption[5]{
	Test of finite-size scaling for the infrared singularities in the 
        dynamical structure factors at $M=1/4$: (a) the longitudinal case 
        at the softmode $p=p_3(M)=\pi/2$. The inset resolves scaling 
        violations for small values of the scaling variable $z_+$; (b) the 
        transverse cases at $p=\pi$.}
\end{figure}

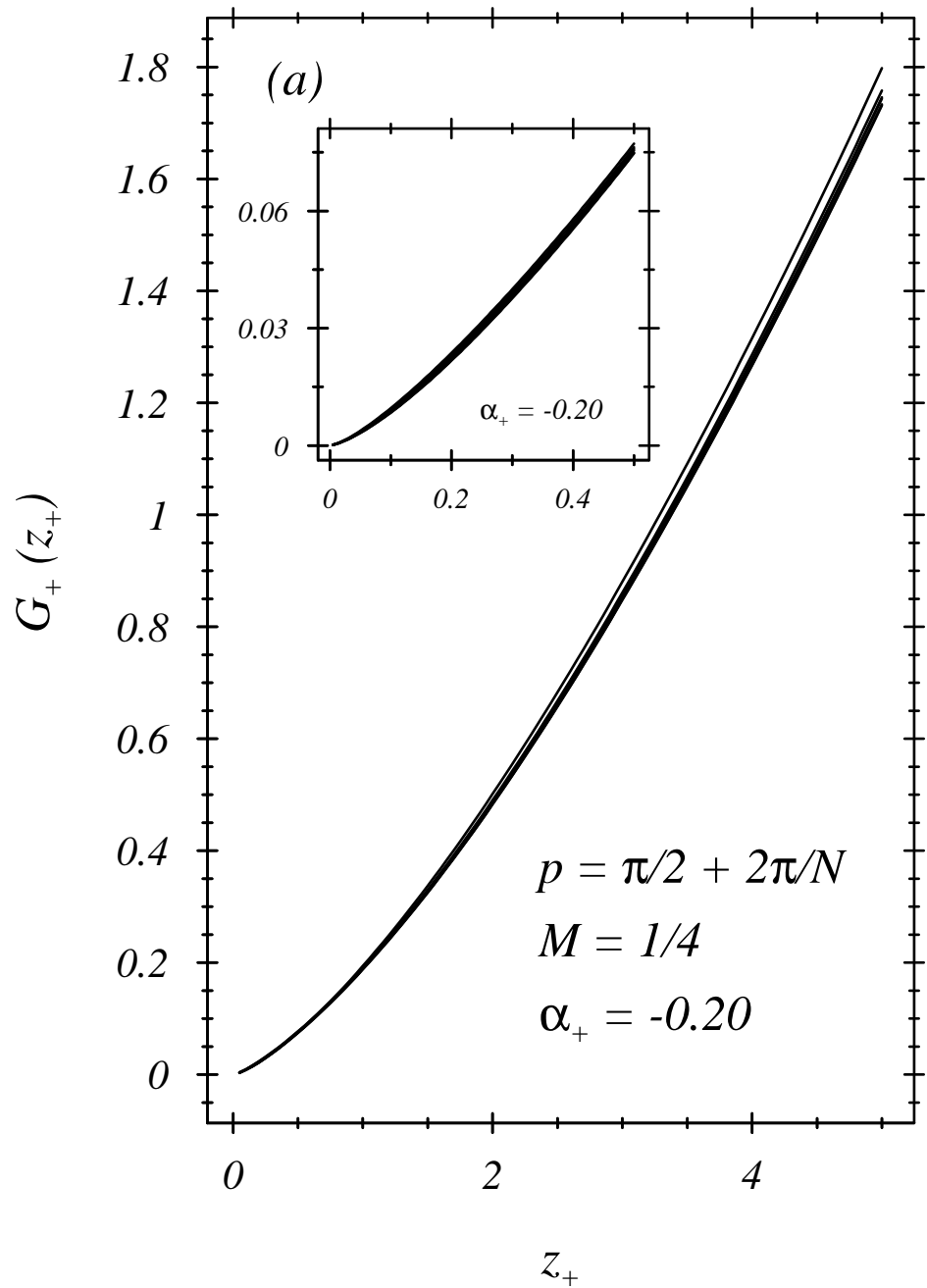
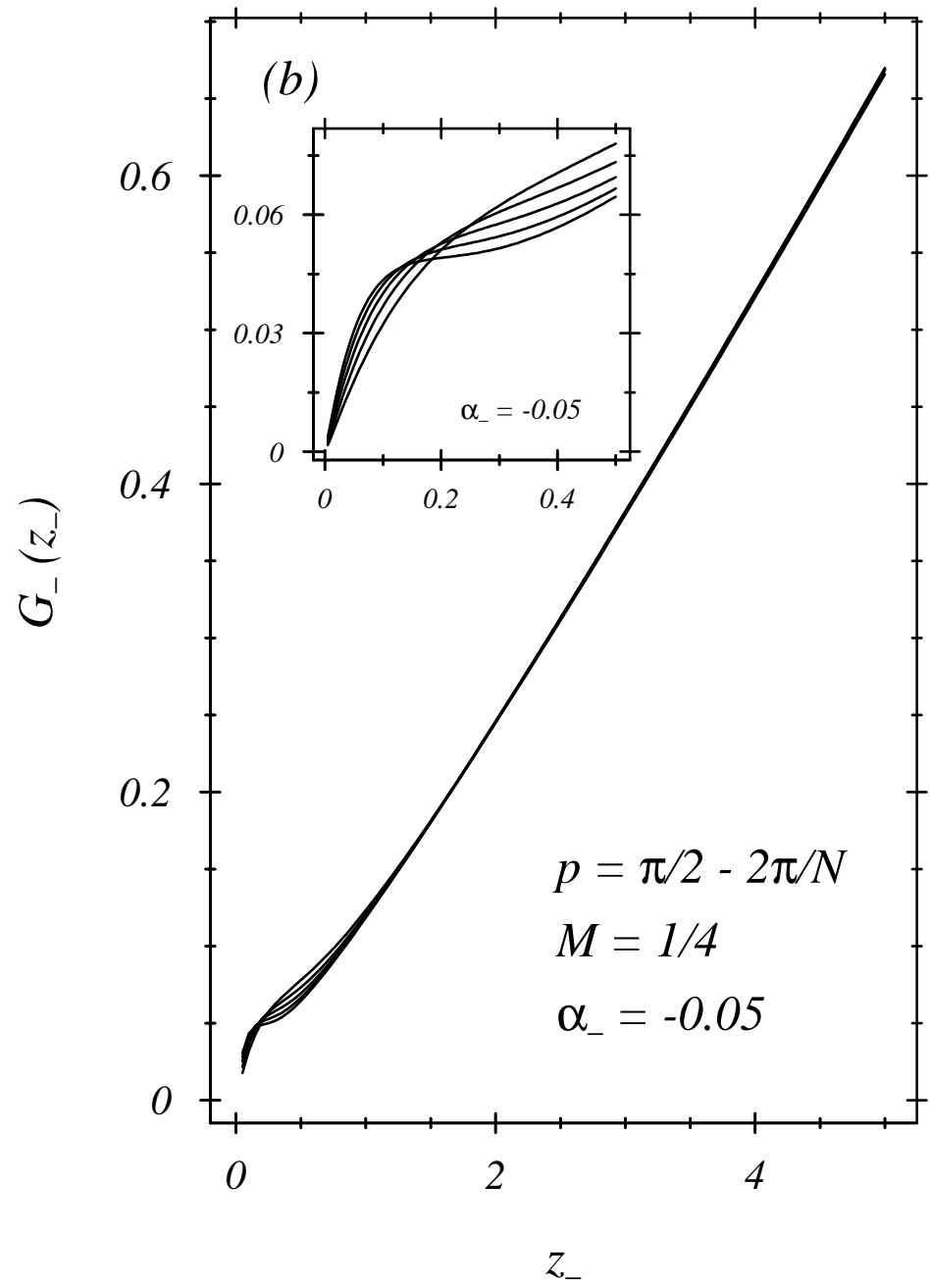
\begin{figure}
\caption[6]{
	Test of finite-size scaling for the infrared singularities in the 
        transverse structure factors at $M=1/4$: (a) the transverse case 
        $S_+$ at the softmode $p=p_1^+(M)=\pi/2+2\pi/N$. The inset shows 
        a magnification for small values of the scaling variable $z_+$; 
        (b) the transverse case $S_-$ at the softmode $p=p_1^-(M)=\pi/2-
        2\pi/N$. The inset resolves scaling violations for small values 
        of the scaling variable $z_-$. }
\end{figure}

\newpage

\begin{table} 
\caption{Energies and transition probabilities for the lowest
excitations in the transverse structure factor $S_-(\omega,p,M,N)$ for $M=1/4$,
$N=16$, $p=\pi$ (left hand part) $p_-=\pi/2-2\pi/16$ (right hand part). The 
upper and lower parts in the table contain the results of an exact 
diagonalization and the recursion method, respectively.}
\begin{minipage}{9cm}
\begin{tabular}{ccc}
\multicolumn{2}{c}{$S_-(\tau=0,p=\pi)=2.52360427892220$} \\ \hline
$\omega_n(\pi)$ & {$w_n(\pi)$} \\ \hline 
0.24490318120407& $7.69543336339913\cdot 10^{-1}$\\
2.00062423661784& $9.81468201828658\cdot 10^{-2}$\\
3.16271478820513& $2.64572507440814\cdot 10^{-4}$\\
3.57865017174411& $6.85304507352309\cdot 10^{-3}$\\
3.98061972078759& $4.71390119166436\cdot 10^{-2}$\\
4.35269652499191& $9.62711159680598\cdot 10^{-5}$\\
4.72994384264668& $8.68880923919877\cdot 10^{-4}$\\
5.11224598930047& $4.94692033004724\cdot 10^{-4}$\\
5.25995835463119& $3.77843015439737\cdot 10^{-5}$\\
5.45318695502460& $7.32076078229036\cdot 10^{-3}$\\
5.74223821730404& $3.48697863610770\cdot 10^{-2}$\\
6.14223417771473& $2.73940664197773\cdot 10^{-6}$\\
6.20371705154730& $2.04948272662971\cdot 10^{-4}$\\
6.28719528119678& $7.30407808674598\cdot 10^{-4}$\\
6.38387404484564& $1.68208400736076\cdot 10^{-5}$\\
6.56406612498208& $1.69023199805961\cdot 10^{-2}$\\
6.76964330648490& $1.81072684633499\cdot 10^{-7}$\\
6.79495897876859& $8.47737682200300\cdot 10^{-3}$\\
6.81533915489440& $3.05825161898085\cdot 10^{-5}$\\
6.83002686340033& $9.95790584921162\cdot 10^{-4}$
\end{tabular}
\end{minipage}
\begin{minipage}{8.5cm}
\begin{tabular}{ccc}
\multicolumn{3}{c}{$S_-(\tau=0,p=p_-)=5.01384876969894\cdot 10^{-1}$} \\ \hline
$\omega_n(p_-)$ & $w_n(p_-)$&\\ \hline
0.87610327625377& $1.95498761012465\cdot 10^{-1}$&*\\
2.50939624323648& $5.59400064383486\cdot 10^{-2}$&\\
3.47398478523209& $6.94575828976292\cdot 10^{-3}$&\\
3.60324922819252& $9.82162858347871\cdot 10^{-4}$&\\
3.71327071028290& $8.01415946306466\cdot 10^{-2}$&\\
4.17033985462645& $7.71455541891509\cdot 10^{-2}$&\\
4.21405414829430& $2.86564726032145\cdot 10^{-4}$&\\
4.30616024321460& $2.50148321929865\cdot 10^{-2}$&\\
4.39960077459270& $1.18063340236613\cdot 10^{-3}$&\\
4.77941756257073& $7.94747197100013\cdot 10^{-3}$&\\
4.99153366093631& $5.41969707921773\cdot 10^{-6}$&\\
5.10045741321637& $5.50730490279844\cdot 10^{-2}$&\\
5.25008778789724& $6.30539985918368\cdot 10^{-2}$&\\
5.37616000377536& $7.35575817722271\cdot 10^{-4}$&\\
5.46963728071208& $1.83331237734851\cdot 10^{-1}$&*\\
5.48768019361761& $1.77004085558857\cdot 10^{-6}$&\\
5.70340946635026& $3.36198337734769\cdot 10^{-6}$&\\
5.71149186560478& $3.44485633165334\cdot 10^{-2}$&\\
5.78088726970425& $1.07071079267298\cdot 10^{-4}$&\\
5.89573570449384& $1.38033053722550\cdot 10^{-1}$&*
\end{tabular}
\end{minipage}
\vspace{0.5cm}

\begin{minipage}{9cm}
\begin{tabular}{ccc}
\multicolumn{2}{c}{$S_-(\tau=0,p=\pi)=2.52360427892349$} \\ \hline 
$\omega_n(p=\pi)$& $w_n(p=\pi)$ \\ \hline
0.24490318120408& $7.69543336339520\cdot 10^{-1}$\\
2.00062423661791& $9.81468201828177\cdot 10^{-2}$\\
3.16271478820483& $2.64572507526320\cdot 10^{-4}$\\
3.57865017173775& $6.85304507238561\cdot 10^{-3}$\\
3.98061972066478& $4.71390118529535\cdot 10^{-2}$\\
4.35269425889094& $9.62693115241292\cdot 10^{-5}$\\
4.72994233277789& $8.68891912517511\cdot 10^{-4}$\\
5.11352567119504& $5.06260574809040\cdot 10^{-4}$\\
5.45159687760725& $7.28553124003179\cdot 10^{-3}$\\
5.74161870216759& $3.48662020479076\cdot 10^{-2}$\\
6.10575246933884& $3.84080214753984\cdot 10^{-4}$\\
6.51195982420624& $1.06776716292014\cdot 10^{-2}$
\end{tabular}
\end{minipage}
\begin{minipage}{8.5cm}
\begin{tabular}{ccc}
\multicolumn{3}{c}{$S_-(\tau=0,p=p_-)=5.01384876970501\cdot 10^{-1}$} \\ \hline
$\omega_n(p_-)$ & $w_n(p_-)$ & \\ \hline
0.87610327625376& $1.95498761012222\cdot 10^{-1}$&*\\
2.50939624323656& $5.59400064382891\cdot 10^{-2}$&\\
3.47398546669344& $6.94594294796492\cdot 10^{-3}$&\\
3.60343461539355& $9.85817067330708\cdot 10^{-4}$&\\
3.71327574076944& $8.01387979043592\cdot 10^{-2}$&\\
4.17072101766960& $7.78964097133079\cdot 10^{-2}$&\\
4.31123417306011& $2.56319236222587\cdot 10^{-2}$&\\
4.77730713351960& $8.29613214115695\cdot 10^{-3}$&\\
5.12926535498599& $8.40410870562240\cdot 10^{-2}$&\\
5.41837248921489& $1.83487171330598\cdot 10^{-1}$&*\\
5.66879948906928& $8.77367998473676\cdot 10^{-2}$&\\
5.94424472415545& $1.41905770467591\cdot 10^{-1}$&*
\end{tabular}
\end{minipage}
\end{table} 

\newpage

\begin{table}
\caption{The critical quantities $2 \theta (M),\;\eta(M)$ and
$2[1-\alpha(M)]$ at $M=1/4$ and at the softmode momenta $p=p_3(M=1/4)=\pi/2$,
$p=p_1^+(M=1/4)$ and $p=p_1^-(M=1/4)$.}
\begin{tabular}{lcccc} 
(a)&$2\theta_3(M)$ & $\eta_3(M)$ & $2[1-\alpha_3(p=\pi/2,M)] $ & \\ 
$p=p_3(M)$ & $1.5312$ & $1.51$ & $1.54$ & \\ \hline
(b)&$2\theta_1(M)$&$\eta_1(M)$&$2[1-\alpha_+(p=\pi,M)]$&$2[1-\alpha_-
(p=\pi,M)]$\\ 
$p=\pi$ & $0.6531$ & $0.65$  & $0.62$ & $0.68$ \\ \hline
(c) & $2\theta_1^+(M)$ & $\eta_1^+(M)$ & $2[1-\alpha_+(p=p_1^+(M),M)]$ & \\ 
$p=p_1^+(M) $ & $2.1843$  &  $2.17$ & $2.40$ & \\ \hline
(d) & $2\theta_1^-(M)$ & $\eta_1^-(M)$ &$2[1-\alpha_-(p=p_1^-(M),M)]$ & \\ 
$p=p_1^-(M) $ & $0.8781$ &  $0.8-1.2$ & $2.1$ & \\ 
\end{tabular} 
\end{table}

\end{document}